# Observation of counterion binding in the inner Helmholtz layer at the ionic surfactant-water interface

Yuyang Peng[1], Feng Gu[1] and Chuanshan Tian[1, a)]

[1]Department of Physics, State Key Laboratory of Surface Physics and Key Laboratory of Micro- and Nano-Photonic Structures (MOE), Fudan University, Shanghai 200433, China

a)e-mail: cstian@fudan.edu.cn

## Abstract

Understanding specific ion adsorption within the inner Helmholtz layer remains central to electrochemistry yet experimentally elusive. Here we directly quantify counterion adsorption and extract the associated thermodynamic parameters within the inner Helmholtz layer using phase-sensitive sum-frequency vibrational spectroscopy (PS-SFVS). Using sodium dodecyl sulfate (SDS) as a model ionic surfactant, we determine the $Na^+$ and $DS^-$ surface densities by simultaneously analyzing interfacial free OH response and the diffuse-layer SF signal, from which the adsorption thermodynamic parameters are derived. We then construct an adsorption phase diagram that maps the evolution of $Na^+$ and $DS^-$ species in the compact layer as functions of bulk NaCl and SDS concentrations, revealing a continuous increase in surface ion pairing. The $DS^-$: $Na^+$ pairing ratio gradually decreases with increasing NaCl and approaches 2.8 at the supersaturation state prior to surface nucleation. These results establish PS-SFVS as a quantitative probe of ion-headgroup correlations in charged interfaces and reveal the thermodynamic mechanism underlying counterion-mediated interfacial ordering, with broad implications for electrolyte design, biomembrane stability, and soft-matter assembly.

# Introduction

At the interface of an electrolyte solution, asymmetric distribution of charges usually leads to the formation of an electrical double layer (EDL). In classical treatments, the EDL is conceptualized as comprising a compact (Helmholtz) region, partitioned into inner and outer Helmholtz planes, and an extended diffuse layer described by Poisson-Boltzmann theory[1]. While this framework, first proposed by Helmholtz[2] in 1853 and later refined by Gouy, Chapman, Stern and others[3–5], is foundational in electrochemistry and colloid science[1,6], it rests on continuum assumptions and leaves open critical questions at the molecular scale: specifically, how ions, dipoles, and polarization fields arrange within the compact Helmholtz region, especially near the inner Helmholtz plane, remains largely inferred rather than directly observed[7,8]. The nanometric thickness and buried nature of this layer pose experimental challenges, limiting our capacity to connect molecular structure to macroscopic interfacial properties.

Ionic surfactants at the air-water interface provide an ideal model system for unraveling the microscopic nature of EDL. Their charged amphiphilic molecules spontaneously assemble into an ordered monolayer, establishing a controllable interfacial electrostatic environment that serves as a benchmark for systematic investigation of the interplay between surface charge, ion adsorption, and molecular structures[9]. This unique interfacial architecture underlies the diverse physicochemical functions of ionic surfactants across applications such as electro-dewetting[10], top-down proteomics[11], and water-based nanofabrication[12]. Extensive macroscopic and scattering studies, including surface tension measurements[9], X-ray scattering[13,14], and neutron reflectivity[15], have shown that Coulombic repulsion among charged headgroups limits molecular packing, while the addition of counterions enhances surface adsorption[9,16] and, at sufficiently high ionic strengths, induces surface crystallization[17]. These observations highlight the delicate balance between surfactant packing and counterion pairing within the inner Helmholtz layer that governs the interfacial phase behavior. Sum-frequency vibrational spectroscopy (SFVS) offers sub-monolayer sensitivity to probe surfactant conformation and interfacial water structure[18–21]. Early SFVS studies showed that NaCl promotes sodium dodecyl sulfate (SDS) adsorption without significantly altering headgroup orientation[22,23]. Heterodyne-detected[24,25] and time-resolved[26] SFVS have further revealed subtle interfacial vibrational signatures associated with organic modes and bonded-OH bands. Nevertheless, the arrangement and influence of counterions in the inner Helmholtz plane remain elusive, hindering our molecular-level understanding of surfactant-mediated interfacial phenomena[21].

In this work, we employ phase-sensitive SFVS (PS-SFVS) to interrogate the aqueous interface of the prototypical ionic surfactant SDS, with a particular focus on the molecular-scale behavior of counterions in the inner Helmholtz layer. By simultaneously analyzing the diffuse-layer SF response and the free OH signal, we directly quantify $Na^+$ accumulation and the enhanced surface density of partially ordered $DS^-$ at the inner Helmholtz layer induced by the addition of NaCl. Through a modified Langmuir adsorption model, we extract a complete set of thermodynamic parameters that characterize the interfacial adsorption equilibria, enabling the construction of an adsorption phase diagram describing the coupled dependencies of $DS^-$ and $Na^+$ surface densities on bulk concentrations. This quantitative framework bridges molecular spectroscopy and thermodynamics, providing a unified picture that explains the onset of surface crystallization as a transition driven by interfacial supersaturation.

# Results and discussion

**Spectroscopic signatures of salt-induced surface crystallization**

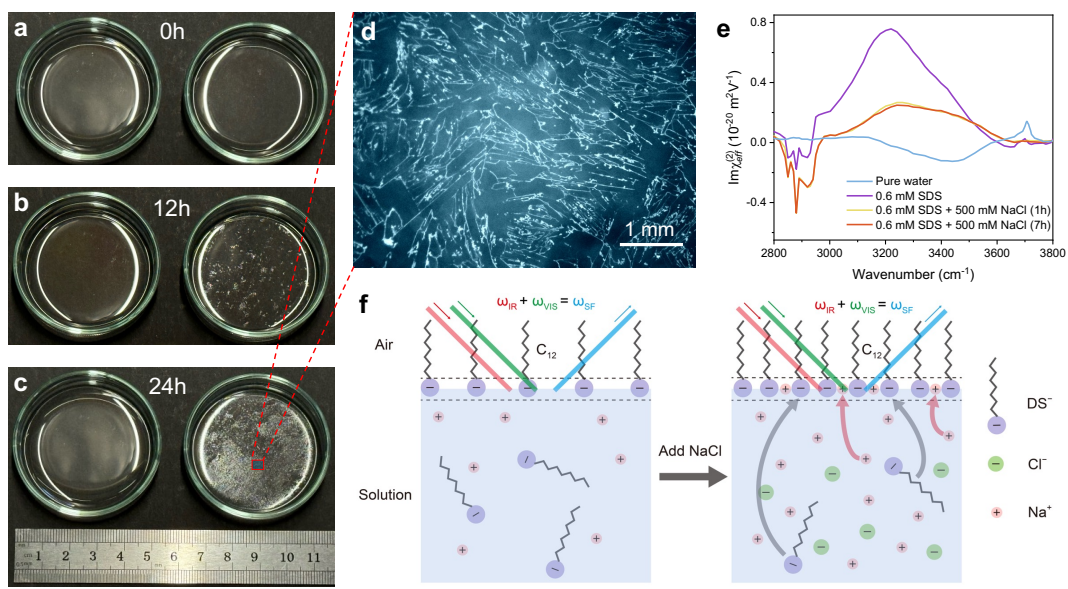

**Fig. 1 | Surface crystallization and interfacial structure of SDS solutions. a-c,** Surface evolution of 0.6 mM SDS (left) and 0.6 mM SDS mixed with 500 mM NaCl (right) in a glass Petri dish at **(a)** 0 h, **(b)** 12 h, and **(c)** 24 h. The room temperature in our experiments is 295 K. **d,** High-magnification image of crystalline surface of the 0.6 mM SDS/500 mM NaCl solution in (c). **e,** $\mathrm{Im}\chi^{(2)}_{S,eff}(\omega)$ spectra at air/water interface of pure water, 0.6 mM SDS solution and 0.6 mM SDS/500 mM NaCl solution at 1 and 7 h. **f,** Schematic illustration of charged interfacial structures: (left) pure SDS solution and (right) SDS solution with NaCl addition.

Figures 1a-c present the time-dependent macroscopic evolution of 0.6 mM SDS solution without (left) and with (right) 500 mM NaCl, demonstrating distinct morphological changes at the interface. In the pure SDS system, the surface morphology remains unchanged over time. In contrast, upon addition of high concentration NaCl, visible crystals nucleate and grow progressively from 8 h, ultimately achieving near-complete surface coverage at 24 h (Fig. 1c). Dark-field microscopic image (Fig. 1d) further highlights these macroscopic crystalline structures. These results confirm prior reports of salt-induced surface crystallization[17].

To reveal the molecular-level structural changes preceding visible crystallization, we employed

PS-SFVS to monitor the interfacial dynamics. Figure 1e shows the SF $\text{Im}\chi^{(2)}_{S,eff}(\omega)$ spectra of air/SDS aqueous interface with and without 500 mM NaCl, using pure water as a reference. In the absence of NaCl, the SF $\text{Im}\chi^{(2)}_{S,eff}(\omega)$ spectrum of 0.6 mM SDS solution exhibits clear C-H vibrational modes in the 2800-3000 cm$^{-1}$ region, indicating adsorption of DS$^-$ at the interface. Meanwhile, a significant decrease of the free-OH peak at 3700 cm$^{-1}$ can be observed owing to hydrogen bond formation between DS$^-$ hydrophilic headgroups and water molecules residing in the topmost layer, which converts the originally free-OH groups into bonded-OH. Additionally, the bonded-OH band in the 3000-3600 cm$^{-1}$ region switches to a positive sign and is enhanced relative to pure water, which is attributed to DS$^-$ adsorption that sets up EDL and orients water molecules with O-H bonds toward the surface[21,27–29].

Upon addition of 500 mM NaCl, the interfacial $\text{Im}\chi^{(2)}_{S,eff}(\omega)$ spectra exhibit significant changes, as shown in Fig. 1e. For the 0.6 mM SDS/500 mM NaCl solution, adding Na$^+$ further increases the C-H vibrational signal, and the free-OH peak now completely vanishes. These changes suggest that DS$^-$ surface coverage approaches a full monolayer as illustrated in Fig. 1f. In the bonded-OH band, on the other hand, one might expect a corresponding increase. However, comparison with the NaCl-free spectrum reveals a decrease, rather than an increase, in the bonded-OH band, which must result from the reduced Debye length after adding NaCl that reduces the amount of polar-oriented water molecules contributing to the SF signal[21,30]. The $\text{Im}\chi^{(2)}_{S,eff}(\omega)$ spectra at 1 h and 7 h show no changes before the emergence of visible crystals, suggesting a stable surface structure until nucleation occurs. It indicates that the interface has reached a critical point of phase transition, i.e. a supercooled state. Therefore, to elucidate the molecular origin of SDS surface aggregation, we performed a quantitative analysis of ion composition and distribution in the Helmholtz layer, particularly within the inner Helmholtz plane, as counterion-specific interactions govern surfactant assembly.

**Quantification of SDS surface density**

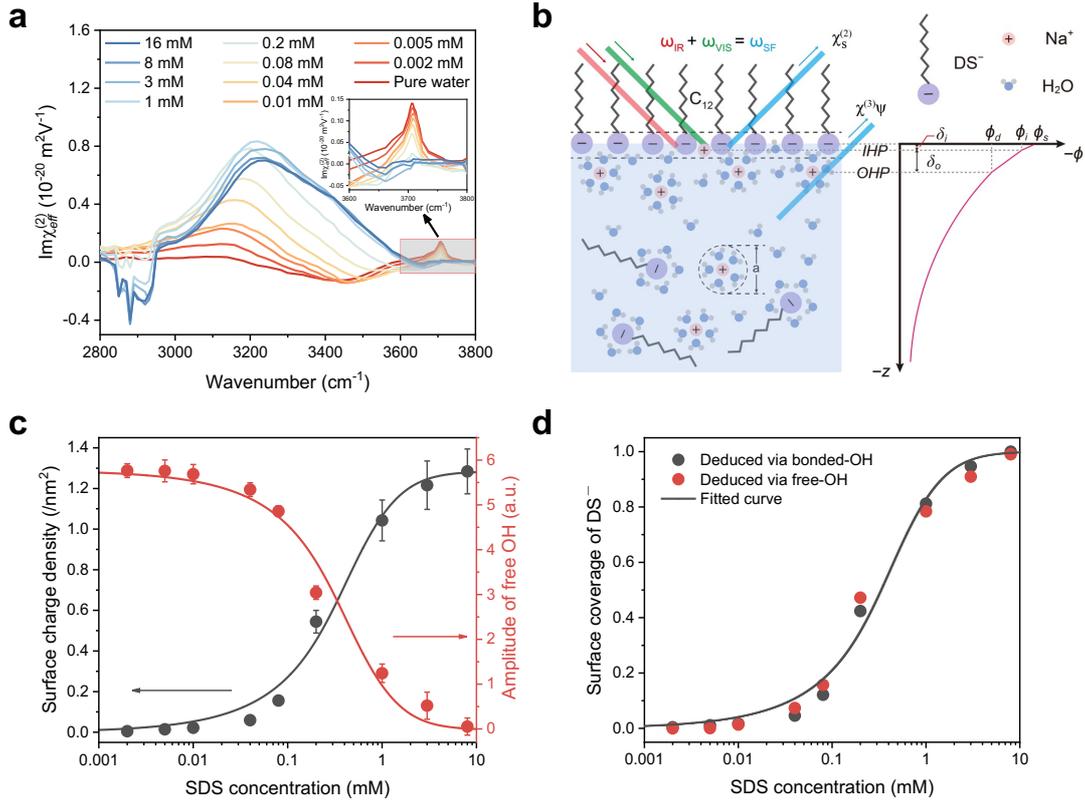

**Fig. 2 | SF spectra and quantitative analysis of pure SDS solutions. a,** $\text{Im}\chi_{S,eff}^{(2)}(\omega)$ spectra at air/water interface of SDS solutions at various concentrations from 0.002 mM to 16 mM. The inset shows a magnified view of the 3600-3800 cm$^{-1}$ region. **b,** Illustration of a charged interfacial structure of pure SDS solution probed by PS-SFVS, $a$ is the effective size of counterions. Two origins contributing to the SF spectrum and the potential distribution in the EDL are also sketched. **c,** Results for the surface charge density (black scatters) and amplitude of free-OH (red scatters) with different bulk SDS concentrations, deduced from modified Gouy-Chapman model and Lorentz peak fitting of free-OH, respectively. The error bars of black dots represent the 95% confidence intervals derived from spectral fitting, while the error bars of red dots represent the standard deviation from the Lorentz peak fitting coefficients. Lines are fitted curves by modified Langmuir adsorption model with $\Delta G_{DS^-} = -38.97 \; kJ/mol$ (see the main text for details). **d,** Normalized data of scatters and fitted curve in **(c)**.

Figure 2a presents the SF spectra of SDS solutions at varying concentrations without NaCl. As SDS concentration increases, the C-H vibrational modes enhance while the free-OH peak decreases, indicating growing surface coverage of DS$^-$. However, the C-H vibrational spectra reveal disordered hydrocarbon chains of DS⁻ as indicated by the gauche defect (2850 cm⁻¹). Thus, it is impossible to quantitatively determine surfactant surface coverage through analysis of C-H vibrations[22,23,31]. We then established a correlation between free-OH depletion and DS⁻ adsorption: adsorption of DS⁻ converts interfacial free OH groups into H-bonded OH, allowing the free-OH signal to serve as a quantitative monitor of DS⁻ surface coverage ($\sigma_{DS^-}$). Figure 2c presents the amplitude of free-OH as a function of SDS concentration (red scatters). At low SDS concentrations (0 to 0.05 mM), it remains hardly changed, meaning minimal adsorption of DS⁻. As concentration increases, free-OH peak rapidly decreases and eventually disappears above 3 mM, suggesting a full monolayer of DS⁻ at the interface.

On the other hand, $\sigma_{DS^-}$ can be independently determined via analysis of spectral change in the H-bonded OH band in Fig. 2a[21]. Strictly speaking, one may obtain the net surface charge density ($\sigma_{net}$) within the inner Helmholtz layer, which is the sum of DS⁻ and Na⁺ densities in this layer ($\sigma_{net} = \sigma_{DS^-} + \sigma_{Na^+}$). As evidenced in Methods and Fig. 4b, for SDS solution without addition of NaCl, $\sigma_{Na^+} \ll \sigma_{DS^-}$ in the inner Helmholtz layer, which leads to $\sigma_{DS^-} \approx \sigma_{net}$. Therefore, $\sigma_{DS^-}$ can be deduced from the spectral change in Fig. 2a using the modified Gouy-Chapman model, which takes into account the finite-size effect of ions, with details given in Methods. The results of calculated $\sigma_{DS^-}$ are given in Fig. 2c (black). After normalization to their respective saturation levels, the DS⁻ surface coverage and the complement of the relative surface coverage of free OH show excellent agreement (Fig. 2d), confirming that the free OH mode reliably quantifies DS⁻ surface coverage.

**Surface excess of counterion with addition of NaCl**

Adding NaCl into SDS solution will result in two consequences: (i) increasing ionic strength and shortening the Debye length; (ii) penetration of counterion Na⁺ into the inner Helmholtz plane and better screening of the repulsion among surface DS⁻. The screening effect reduces the surface energy and promotes more DS⁻ adsorbing at the surface. Fig. 3a presents SF spectra of 0.01 mM SDS solutions with various concentrations of NaCl. The intensity of C-H stretching vibrational modes (2800-3000 cm⁻¹) increases monotonically with NaCl concentration while the intensity of free-OH decreases consistently, indicating increased surface coverage of hydrocarbon chains. The H-bonded OH band initially increases with DS⁻ adsorption, but then decreases when NaCl concentration exceeds 1 mM as the reduced Debye length suppresses the diffuse-layer contribution to the SF signal[21,32].

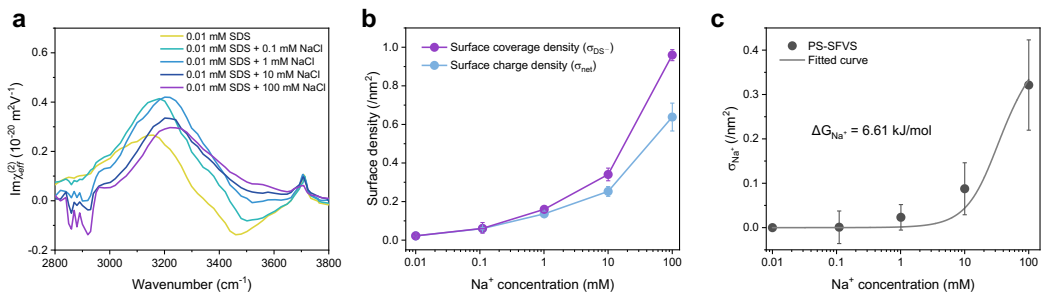

**Fig. 3 | SF spectra and quantitative analysis for Na⁺ surface excess. a,** $\mathrm{Im}\chi^{(2)}_{S,eff}(\omega)$ spectra at air/water interface of 0.01 mM SDS solutions with various NaCl concentrations from 0.1 mM to 100 mM. **b,** Results for the surface charge density and surface coverage density with different NaCl concentrations, deducing from two different methods (see the main text for details). Lines are guides to the eyes. **c,** Deduced Na⁺ surface densities versus bulk Na⁺ concentration in 0.01 mM SDS solutions, fitted by ion adsorption isotherm.

For quantitative analysis, the surface charge density ($\sigma_{net}$) at different NaCl concentrations can be deduced by quantitative analysis of the 3000-3600 cm⁻¹ region, while the surface coverage density ($\sigma_{DS^-}$) can be interpolated from free-OH amplitude using the data in Fig. 2c. The comparison between the $\sigma_{net}$ and $\sigma_{DS^-}$ is displayed in Fig. 3b. As expected, at NaCl concentrations less than 1 mM, $\sigma_{net} \approx \sigma_{DS^-}$. Notably, with more NaCl addition, $\sigma_{DS^-}$ progressively exceeds the net surface charge density. These findings provide definitive experimental evidence that the dehydrated Na⁺ ions must have penetrated into the inner Helmholtz layer and partially neutralize interfacial charge via forming "counterion bridges" between charged headgroups of surfactants. The deduced Na⁺ surface densities ($\sigma_{Na^+} = \sigma_{net} - \sigma_{DS^-}$), shown in Fig. 3c, remain negligible for bulk Na⁺ concentrations below 1 mM and increase to ~30% of the DS⁻ surface coverage at 100 mM. Thus, compared to pure SDS solutions, SDS solutions with NaCl exhibit surface excess of Na⁺. This surface structure exerts a shielding effect on the Coulombic repulsion between surfactant headgroups, thereby facilitating enhanced DS⁻ adsorption at the interface even under low bulk SDS concentrations. This shielding mechanism promotes greater surfactant accumulation at the surface, leading to significantly increased surface coverage beyond conventional limits.

**Adsorption kinetics at the charged interface**

To further quantify the molecular adsorption kinetic process at the interface[33], we consider two equilibrium reactions for the adsorption of $DS^-$ and $Na^+$

$$DS^-_{bulk} \overset{K_S}{\leftrightarrow} DS^-_{surface} \tag{1}$$

$$DS^-_{surface} + Na^+_{bulk} \overset{K_{Na}}{\leftrightarrow} DS^-Na^+_{surface} \tag{2}$$

where $K_S = exp\left(-\frac{\Delta G^0_S}{RT}\right)$ and $K_{Na} = exp\left(-\frac{\Delta G^0_{Na}}{RT}\right)$ are the corresponding equilibrium constants for the surface adsorption, $\Delta G^0_S$ and $\Delta G^0_{Na}$ are the standard Gibbs free energy. However, for ionic surfactant solutions, the adsorption at the interface must consider two competing interactions: the attractive van der Waals interactions between hydrocarbon tails and the Coulomb repulsion between charged headgroups (quantified by the energy associated with the surface potential), which are expressed as $exp\left(\frac{2\beta\theta_S}{k_BT}\right)$ and $exp\left(\frac{e\phi_s}{k_BT}\right)$, respectively[34,35]. Here $\phi_s$ is the surface potential accounting for voltage drop across the Stern layer, which takes the form[1]

$$\phi_i = \phi_d + \frac{\delta_o e \sigma_{net}}{\varepsilon_0 \varepsilon_s} \tag{3}$$

$$\phi_s = \phi_i + \frac{\delta_i e \sigma_{DS^-}}{\varepsilon_0 \varepsilon_s} \tag{4}$$

As shown in Fig. 2b, the Helmholtz layer is modeled as a parallel capacitor defined by its thickness $\delta_i+\delta_o$ and the interfacial relative dielectric constant $\varepsilon_s$ in water, here we adopt $\delta_o = a/2 = 0.35\ nm$[21,36,37] and $\varepsilon_s = 6$ [38,39]. According to molecular dynamics simulations[40], we set $\delta_i \approx 0\ nm$ under the assumption that the counterions can adsorb at the same plane as the surfactant headgroups[41]. In practice, the results presented in Extended Data Fig. 1 demonstrate that varying $\delta_i$ within a range of 0 to 0.2 nm[37,42] does not significantly affect the quality of the curve fitting.

Accounting for the adsorption processes of $Na^+$ and $DS^-$ and the interactions between these ions, we adopt the modified Langmuir adsorption equations[34,35]

$$K_S \cdot exp\left(\frac{e\phi_s}{k_BT}\right) exp\left(\frac{2\beta\theta_S}{k_BT}\right) \cdot \frac{C_{SDS}}{C_S} = \frac{\theta_S}{1-\theta_S} \quad (5)$$

$$K_{Na} \cdot exp\left(-\frac{e\phi_i}{k_BT}\right) \cdot \frac{C_{Na^+}}{C_S} = \frac{\theta_{Na}}{1-\theta_{Na}} \quad (6)$$

where $\theta_S = \frac{\sigma_{DS^-}}{\sigma_\infty}$ and $\theta_{Na} = -\frac{\sigma_{Na^+}}{\sigma_{DS^-}}$ are the surface fractional coverage of DS⁻ and Na⁺, respectively, $C_{SDS}$ and $C_{Na^+}$ are the bulk concentrations, $C_S = 1 \ mol/L$ is the standard concentration[1,43], and $\beta$ is the Frumkin interaction parameter[34,35], which is positive for attractive interactions. On this basis, we fitted the data in Fig. 2d and Fig. 3c with equations (5) and (6), obtaining the standard Gibbs free energy of adsorption $\Delta G_S^0 = -38.97 \pm 1.18$ kJ/mol, $\Delta G_{Na}^0 = 6.61 \pm 0.82$ kJ/mol and $2\beta/k_BT = 3.2 \pm 1.0$, respectively. The comparison of surface charge densities fitted using different $\beta$ values is provided in Extended Data Fig. 2, the value of $\beta$ obtained from our fitting is consistent with that reported in ref.[44]. Notably, after isolating the Coulomb attraction term from the equilibrium constant, the derived adsorption free energy of Na⁺ is positive, reflecting its hydration-shell stabilization that disfavors dehydration for adsorption, which is a clear indication that Na⁺ adsorption is primarily driven by Coulomb interactions.

**Thermodynamic and structural evolution at the interface**

Based on theoretical modeling and experimental analysis described above, we have successfully extracted a complete set of thermodynamic parameters that characterize ion adsorption within the compact layer. These parameters allow quantitative prediction of the surface densities of DS⁻ and Na⁺, as well as their stoichiometric ratio within the inner Helmholtz layer, for any given combination of bulk ion concentrations (Fig. 4a, b). The grey shaded regions represent concentration combinations that cannot be physically achieved, with their boundaries defined by the pure SDS solution. As the concentrations of DS⁻ (vertical axis) or Na⁺ (horizontal axis) increase, more ionic surfactant appears on the surface while the Na⁺ counterions are also attracted to the surface as evidenced by the decreased DS⁻-Na⁺ ratio. In Fig. 4a, the three dashed lines represent 25%, 50%, and 75% surface coverage of SDS, respectively. Correspondingly, Fig. 4b shows dashed lines indicating Na⁺ surface densities equivalent to 0.1%, 1%, 10% and 25% of the DS⁻ in the inner Helmholtz layer. For the represented line cuts at 0.6 mM of DS⁻, Fig. 4c shows the surface density of DS⁻ and the ratio of DS⁻-Na⁺ in 0.6 mM SDS solutions with various Na⁺ concentrations.

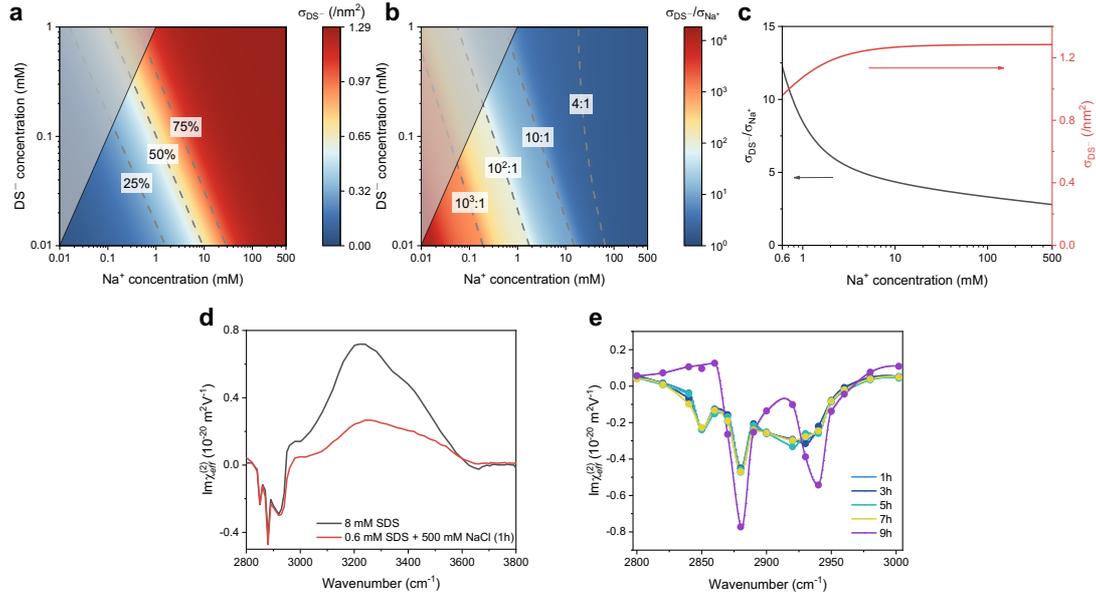

**Fig. 4 | Thermodynamic modeling and time-dependence SF spectra of SDS-NaCl solutions. a, b,** Calculated surface density of **(a)** DS⁻ and **(b)** ratio of DS⁻-Na⁺ in the inner Helmholtz layer using the modified Langmuir adsorption model for various bulk concentrations of DS⁻ and Na⁺. The grey shaded area represents non-physically existing concentration combinations ($C_{SDS} < C_{Na^+}$), and dashed lines indicates major levels of contour lines. **c,** Represented line cuts in **(a)** and **(b)** with the bulk concentration of SDS set as 0.6 mM. **d,** $\text{Im}\chi^{(2)}_{S,eff}(\omega)$ spectra at air/water interface of 8 mM SDS solution and 0.6 mM SDS/500 mM NaCl solution. **e,** Time-dependence of $\text{Im}\chi^{(2)}_{S,eff}(\omega)$ spectra of 0.6 mM SDS/500 mM NaCl solution in the 2800-3000 cm⁻¹ region.

This quantitative description enables a molecular-level understanding of the surface crystallization process observed in Fig. 1. As shown by the curves in Fig. 4c, addition of NaCl in pure SDS solution increases DS⁻ surface density from 0.9 nm⁻² to 1.3 nm⁻². Meanwhile, the Na⁺ surface concentration grows until $\sigma_{DS^-}/\sigma_{Na^+} \approx 2.8$, suggesting near-saturation of counterion binding in the inner Helmholtz layer. On the other hand, addition of 500 mM NaCl perturbs the bulk equilibrium of SDS, reducing the critical micelle concentration (CMC) to 0.5 mM and lowering the Krafft point to 25°C. Under these conditions, 0.6 mM SDS at room temperature falls below the shifted Krafft point and exceeds the depressed CMC, reaching spontaneous crystallization instead of micelle formation. The accumulation of DS⁻ and Na⁺ at the surface thus creates a supersaturated interfacial state, where partially dehydrated Na⁺ ions in the inner Helmholtz layer reduce the nucleation barrier and trigger surface crystallization.

To confirm the supersaturated state on the surface before crystallization, we compared SF spectra of two representative systems: (i) 8 mM SDS solution that has reached CMC and (ii) 0.6 mM SDS solution containing 500 mM NaCl. The results are presented in Fig. 4d, where the two SF spectra are identical in the CH and free OH region. (The difference in the H-bonded OH band arises from different ionic strength (Debye length).) It is known that, at the CMC, the surface coverage of $DS^-$ reaches its maximum. Therefore, the solution of 0.6 mM SDS with 500 mM NaCl has also reached saturation on the surface. The time-dependent spectra of 0.6 mM SDS after adding 500 mM NaCl, shown in Fig. 4e, also confirm this conclusion, where the spectrum remains identical up to ~8 h until the visible particulate crystallites emerge at 9 h, consistent with delayed nucleation from a supersaturated interface. It is worthwhile to note that, even at maximum surface coverage, the $DS^-$ molecules are not densely packed as evidenced by the persistence of the symmetric stretching mode at 2850 $cm^{-1}$ due to gauche defects. However, once crystallized on surface, the $CH_2$ peak (2850 $cm^{-1}$) diminishes while the $CH_3$ peak increases (purple dots for 9 h in Fig. 4e), indicating a disorder-to-order transition of SDS tails.

## Conclusion

In summary, this study establishes a quantitative thermodynamic framework for understanding ion-specific adsorption in the inner Helmholtz layer of charged aqueous interfaces. Using PS-SFVS, we directly quantified the interfacial $Na^+$ and $DS^-$ surface densities, from which the adsorption thermodynamic parameters were extracted. These thermodynamic parameters enabled the construction of an adsorption phase diagram that delineates the coupled evolution of $Na^+$ and $DS^-$ populations as a function of SDS and NaCl concentrations, revealing a continuous strengthening of ion-pairing within the compact layer. The gradual decrease of the $DS^-$ : $Na^+$ ratio toward 2.8 near the supersaturation limit marks the onset of surface nucleation driven by ion–headgroup correlation. This quantitative mapping of interfacial thermodynamics provides a molecular-level understanding of how counterion-mediated interactions govern surfactant adsorption, aggregation, and crystallization. The approach offers a generalizable route for probing ion-specific phenomena at soft charged interfaces, with broad implications for electrolyte design, biomembrane stability, and the controlled assembly of functional nanostructures.

# Methods

**Sample preparation**

SDS (BioXtra, ≥ 99.0%, GC) and NaCl (Suprapur, ≥ 99.99%) were purchased from Sigma-Aldrich, and all the solutions were prepared in ultrapure water (18.2 MΩ·cm, Thermo fisher scientific). Before every experiment, the SDS powder was recrystallized and vacuum-dried at least three times from ethanol to remove the dodecanol impurities[22], which are known to be the hydrolysis product of SDS[26,46,47]. Extended Data Fig. 3 shows the SF spectra with and without further purification. For SDS solutions near CMC, the presence of dodecanol can highly affect the surface density of molecules without changing the charge density at the interface, due to its higher surface propensity than SDS. However, when the concentration exceeds the CMC, dodecanol molecules preferentially localize at micelle interface rather than the air-water interface[46]. Thus, the nearly identical surface organic vibrational peaks between purified 8 mM and 16 mM SDS solutions can confirm successful impurity removal (Extended Data Fig. 3b). Notably, the fused silica cell and other tools in contact with sample must be very well cleaned. They were first soaked in a mixture solution (~4 g/100 mL) of sulfuric acid (98%, Titan) and NOCHROMIX (Godax laboratories) for at least 16 hours, then rinsed with ultrapure water (18.2 MΩ·cm, Thermo fisher scientific) and dried by a constant-temperature heating platform at 120 ˚C.

**Phase-sensitive sum-frequency vibrational spectroscopy measurement**

Our PS-SFVS setup has been described in detail previously[29]. We used a 30-ps, 20-Hz Nd: YAG pulsed laser (PL2551-A, Ekspla) and a home-built THG/OPA/DFG system to generate a s-polarized visible beam at 532 nm and a p-polarized tunable IR beam which can cover a range of 2800 to 3800 cm$^{-1}$ with enough pulse energy for SFVS. Then the two beams were collinearly focused and overlapped on the sample at 45˚ with pulse energy/spot diameter of 300 µJ/1.4 mm and 50 µJ/1.3 mm, respectively. The coherent length of this geometry is ~30 nm. For the phase measurement of SF signal, we inserted a 50 µm y-cut quartz and a fused silica plate as local oscillator (LO) and phase modulator (PM), respectively. The s-polarized SF signal generated from the LO interfered with that from the sample in reflection, and a SF interferogram can be obtained by rotating the PM. Then $Im\chi^{(2)}_{S,eff}(\omega)$ was deduced from the measured amplitude and phase of SF output signal from the sample, which were normalized to that from a z-cut alpha-quartz. Note that all spectra in this paper have been Fresnel-factor corrected.

**Modified Gouy-Chapman model**

According to the established theory[21], the SF responses from the EDL can be expressed as

$$\chi^{(2)}_{S,eff}(\omega) = \chi^{(2)}_S(\omega) + \chi^{(3)}(\omega) \cdot \Psi$$

$$= \chi^{(2)}_S(\omega) + \chi^{(3)}(\omega) \cdot \int_{0^+}^{\infty} \hat{z}\, E_0(z) e^{i\Delta k_z z} dz \quad (7)$$

where $\Delta k_z = k_{VIS,z} + k_{IR,z} + k_{SF,z}$ is the phase mismatch in reflection geometry, $E_0(z)$ is the DC electric field in the diffuse layer, and $\chi^{(2)}_S(\omega)$ and $\chi^{(3)}(\omega)$ are the surface and bulk nonlinear susceptibilities of water, respectively. The bulk $\chi^{(3)}(\omega)$ spectrum can be measured independently. The integral in the $\chi^{(3)}(\omega)$ contribution is related to the electric field distribution in the EDL, which is dictated by surface charge and ionic strength. Thus, when $\chi^{(2)}_S(\omega)$ is known a priori or its change is negligible versus the diffuse layer response[32], we can obtain the surface charge density $\sigma_{net}$ by analyzing the spectral changes.

We first focus on $\sigma_{net}$ by quantifying changes in the bonded-OH band. A simple derivation from equation (7) yields

$$Im\chi^{(2)}_{S,eff}(\omega) = Im\chi^{(2)}_S(\omega) + Im\chi^{(3)}(\omega) \cdot \int_{0^+}^{\infty} \hat{z}\, E_0(z) cos(\Delta k_z z) dz$$

$$+ Re\chi^{(3)}(\omega) \cdot \int_{0^+}^{\infty} \hat{z}\, E_0(z) sin(\Delta k_z z) dz \quad (8)$$

Note that when the Debye length is larger than or comparable to the coherent length (~ 30 nm), e.g., at ionic strength < 1 mM, the $Re\chi^{(3)}(\omega)$ contribution significantly influences the spectral lineshape; else, it is negligible[21,32]. Here, we adopt the modified Gouy-Chapman model to calculate the DC electric field distribution $E_0(z)$, which takes into account the finite-size effect of ions (Fig. 2b). The modified Poisson-Boltzmann equation takes the form[48,49]

$$\nabla^2 \phi(z) = \frac{eC}{\epsilon_0 \epsilon_r} \frac{2 \sinh\left(\frac{e\phi}{k_B T}\right)}{1 + 2v \sinh^2\left(\frac{e\phi}{2k_B T}\right)} \tag{9}$$

and

$$\phi_d(z = 0^+) = -\frac{2k_B T}{e} \sinh^{-1}\left\{\sqrt{\frac{1}{2v}\left[\exp\left(\frac{ve^2 \sigma_{net}^2}{4C\epsilon_0 \epsilon_r k_B T}\right) - 1\right]}\right\} \tag{10}$$

Here, $v = 2a^3 C$ is the ion packing parameter, with $a$ being the effective size of hydrated counterions and $C$ being the ionic strength. In Extended Data Fig. 4, we have compared surface charge densities calculated using different $a$ values. For SDS concentrations below 0.1 mM, the effect of $a$ is no longer significant. In the following calculation, we adopt $a$ = 0.7 nm for hydrated $Na^+$ as reported in literature[21,36,37,48,49].

Based on the preceding description, the $\chi_S^{(2)}(\omega)$ spectrum of SDS solution can be effectively approximated as equivalent to that of pure water[32]. Therefore, surface charge density $\sigma$ is the only unknown variable in equations (8)-(10). By comparing calculated $Im\chi_{S,eff}^{(2)}(\omega)$ spectra (generated using different $\sigma$ values) with experimental spectra via least-squares fitting, the surface charge density $\sigma$ (i.e., the surface density of $DS^-$) is determined. The calculated spectra and experimental spectrum are shown in Extended Data Fig. 5, indicating that $\sigma = 0.0054 \pm 0.0002\ e/nm^2$ in 0.002 mM SDS solution. Extended Data Fig. 6b compares the surface density of $DS^-$ obtained via PS-SFVS with that derived from surface tension measurements[50]. The results from both methods align well.

**Surface $Na^+$ adsorption in pure SDS solution**

Considering that pure SDS solutions also contain a small amount of dehydrated $Na^+$ adsorbed at the interface, the data in Fig. 3b essentially compare the surface charge densities between two cases at identical surface coverage: (1) high-concentration pure SDS (Extended Data Fig. 7a) and (2) low-concentration SDS with added NaCl (Extended Data Fig. 7b). By applying results in Fig. 3b to equation (6) and utilizing the constant nature of $K_{Na}$, we can determine the respective surface charge densities of adsorbed $Na^+$ for both cases (Extended Data Fig. 7c). The results reveal that in pure dilute SDS solutions, $Na^+$ predominantly remains hydrated in the Stern layer, with only minimal dehydrated Na$^+$ adsorbed in the inner Helmholtz layer. Consequently, the surface charge density obtained from PS-SFVS follows $\sigma_{net1} = \sigma_{DS^-}$ for pure SDS solutions, while for NaCl-added SDS solutions it becomes $\sigma_{net2} = \sigma_{DS^-} + \sigma_{Na^+}$.

# References

1. Bard, A. J. & Faulkner, L. R. *Electrochemical Methods: Fundamentals and Applications*. (Wiley, New York Weinheim, 2001).
2. Helmholtz, H. Ueber einige Gesetze der Vertheilung elektrischer Ströme in körperlichen Leitern mit Anwendung auf die thierisch-elektrischen Versuche. *Annalen der Physik* **165**, 211–233 (1853).
3. Gouy, M. Sur la constitution de la charge électrique à la surface d'un électrolyte. *J. Phys. Theor. Appl.* **9**, 457–468 (1910).
4. Chapman, D. L. A contribution to the theory of electrocapillarity. *The London, Edinburgh, and Dublin Philosophical Magazine and Journal of Science* **25**, 475–481 (1913).
5. Stern, O. ZUR THEORIE DER ELEKTROLYTISCHEN DOPPELSCHICHT. *Zeitschrift für Elektrochemie und angewandte physikalische Chemie* **30**, 508–516 (1924).
6. Lyklema, H. *Fundamentals of Interface and Colloid Science*. (Academic Press, San Diego, 2000).
7. Magnussen, O. M. & Groß, A. Toward an Atomic-Scale Understanding of Electrochemical Interface Structure and Dynamics. *J. Am. Chem. Soc.* **141**, 4777–4790 (2019).
8. Le, J.-B., Fan, Q.-Y., Li, J.-Q. & Cheng, J. Molecular origin of negative component of Helmholtz capacitance at electrified Pt(111)/water interface. *Sci. Adv.* **6**, eabb1219 (2020).
9. Rosen, M. J. & Kunjappu, J. T. *Surfactants and Interfacial Phenomena*. (Wiley, Hoboken, N.J, 2012).
10. Li, J., Ha, N. S., Liu, T. 'Leo', Van Dam, R. M. & 'Cj' Kim, C.-J. Ionic-surfactant-mediated electro-dewetting for digital microfluidics. *Nature* **572**, 507–510 (2019).
11. Brown, K. A. *et al.* A photocleavable surfactant for top-down proteomics. *Nat Methods* **16**, 417–420 (2019).
12. Xie, W. *et al.* Solvent-pair surfactants enabled assembly of clusters and copolymers towards programmed mesoporous metal oxides. *Nat Commun* **14**, 8493 (2023).
13. Bloch, J. M. *et al.* Concentration Profile of a Dissolved Polymer near the Air-Liquid Interface: X-Ray Fluorescence Study. *Phys. Rev. Lett.* **54**, 1039–1042 (1985).
14. Kuhl, T. L. *et al.* Packing Stress Relaxation in Polymer−Lipid Monolayers at the Air−Water Interface: An X-ray Grazing-Incidence Diffraction and Reflectivity Study. *J. Am. Chem. Soc.* **121**, 7682–7688 (1999).
15. Lu, J. R., Hromadova, M., Simister, E. A., Thomas, R. K. & Penfold, J. Neutron Reflection from Hexadecyltrimethylammonium Bromide Adsorbed at the Air/Liquid Interface: The Variation of the Hydrocarbon Chain Distribution with Surface Concentration. *J. Phys. Chem.* **98**, 11519–11526 (1994).
16. Judd, K. D., Parsons, S. W., Majumder, T. & Dawlaty, J. M. Electrostatics, Hydration, and Chemical Equilibria at Charged Monolayers on Water. *Chem. Rev.* **125**, 2440–2473 (2025).

17. Kharlamova, A. *et al.* Interface-Templated Crystal Growth in Sodium Dodecyl Sulfate Solutions with NaCl. *Langmuir* **40**, 84–90 (2024).

18. Shen, Y. R. Surface properties probed by second-harmonic and sum-frequency generation. *Nature* **337**, 519–525 (1989).

19. Ji, N., Ostroverkhov, V., Tian, C. S. & Shen, Y. R. Characterization of Vibrational Resonances of Water-Vapor Interfaces by Phase-Sensitive Sum-Frequency Spectroscopy. *Phys. Rev. Lett.* **100**, 096102 (2008).

20. Tian, C. S. & Shen, Y. R. Structure and charging of hydrophobic material/water interfaces studied by phase-sensitive sum-frequency vibrational spectroscopy. *Proc. Natl. Acad. Sci. U.S.A.* **106**, 15148–15153 (2009).

21. Wen, Y.-C. *et al.* Unveiling Microscopic Structures of Charged Water Interfaces by Surface-Specific Vibrational Spectroscopy. *Phys. Rev. Lett.* **116**, 016101 (2016).

22. Johnson, C. M. & Tyrode, E. Study of the adsorption of sodium dodecyl sulfate (SDS) at the air/water interface: targeting the sulfate headgroup using vibrational sum frequency spectroscopy. *Phys. Chem. Chem. Phys.* **7**, 2635 (2005).

23. Hore, D. K., Beaman, D. K. & Richmond, G. L. Surfactant Headgroup Orientation at the Air/Water Interface. *J. Am. Chem. Soc.* **127**, 9356–9357 (2005).

24. Nihonyanagi, S., Yamaguchi, S. & Tahara, T. Water Hydrogen Bond Structure near Highly Charged Interfaces Is Not Like Ice. *J. Am. Chem. Soc.* **132**, 6867–6869 (2010).

25. Sengupta, S., Versluis, J. & Bakker, H. J. Observation of a Two-Dimensional Hydrophobic Collapse at the Surface of Water Using Heterodyne-Detected Surface Sum-Frequency Generation. *J. Phys. Chem. Lett.* **14**, 9285–9290 (2023).

26. Livingstone, R. A., Nagata, Y., Bonn, M. & Backus, E. H. G. Two Types of Water at the Water–Surfactant Interface Revealed by Time-Resolved Vibrational Spectroscopy. *J. Am. Chem. Soc.* **137**, 14912–14919 (2015).

27. Miranda, P. B., Du, Q. & Shen, Y. R. Interaction of water with a fatty acid Langmuir film. *Chemical Physics Letters* **286**, 1–8 (1998).

28. Mondal, J. A., Nihonyanagi, S., Yamaguchi, S. & Tahara, T. Three Distinct Water Structures at a Zwitterionic Lipid/Water Interface Revealed by Heterodyne-Detected Vibrational Sum Frequency Generation. *J. Am. Chem. Soc.* **134**, 7842–7850 (2012).

29. Xu, Y., Ma, Y.-B., Gu, F., Yang, S.-S. & Tian, C.-S. Structure evolution at the gate-tunable suspended graphene–water interface. *Nature* **621**, 506–510 (2023).

30. Hunger, J. *et al.* Nature of Cations Critically Affects Water at the Negatively Charged Silica Interface. *J. Am. Chem. Soc.* **144**, 19726–19738 (2022).

31. Zhuang, X., Miranda, P. B., Kim, D. & Shen, Y. R. Mapping molecular orientation and conformation at interfaces by surface nonlinear optics. *Phys. Rev. B* **59**, 12632–12640 (1999).

32. Hsiao, Y., Chou, T.-H., Patra, A. & Wen, Y.-C. Momentum-dependent sum-frequency vibrational spectroscopy of bonded interface layer at charged water interfaces. *Sci. Adv.* **9**, eadg2823 (2023).

33. Langmuir, I. The adsorption of gases on plane surfaces of glass, mica, and platinum. *J. Am. Chem. Soc.* **40**, 1361–1403 (1918).

34. Koryta, J., Dvořák, J. & Kavan, L. *Principles of Electrochemistry*. (Wiley, Chichester, 1993).

35. *Soft Matter at Aqueous Interfaces*. vol. 917 (Springer International Publishing, Cham, 2016).

36. Lyubartsev, A. P. & Laaksonen, A. Concentration Effects in Aqueous NaCl Solutions. A

Molecular Dynamics Simulation. *J. Phys. Chem.* **100**, 16410–16418 (1996).

37. Marcus, Y. Ionic radii in aqueous solutions. *Chem. Rev.* **88**, 1475–1498 (1988).
38. Chen, F., Qing, Q., Xia, J., Li, J. & Tao, N. Electrochemical Gate-Controlled Charge Transport in Graphene in Ionic Liquid and Aqueous Solution. *J. Am. Chem. Soc.* **131**, 9908–9909 (2009).
39. Dinpajooh, M. & Matyushov, D. V. Dielectric constant of water in the interface. *The Journal of Chemical Physics* **145**, (2016).
40. Tiwari, S. Effect of salt on the adsorption of ionic surfactants at the air-water interface. *Journal of Molecular Liquids* (2022).
41. Warszyński, P., Barzyk, W., Lunkenheimer, K. & Fruhner, H. Surface Tension and Surface Potential of Na *n*-Dodecyl Sulfate at the Air−Solution Interface: Model and Experiment. *J. Phys. Chem. B* **102**, 10948–10957 (1998).
42. Shannon, R. D. Revised effective ionic radii and systematic studies of interatomic distances in halides and chalcogenides. *Acta Cryst A* **32**, 751–767 (1976).
43. Liu, Y. Is the Free Energy Change of Adsorption Correctly Calculated? *J. Chem. Eng. Data* **54**, 1981–1985 (2009).
44. Prosser, A. J. & Franses, E. I. Adsorption and surface tension of ionic surfactants at the air–water interface: review and evaluation of equilibrium models. *Colloids and Surfaces A: Physicochemical and Engineering Aspects* **178**, 1–40 (2001).
45. Phillips, J. N. The energetics of micelle formation. *Trans. Faraday Soc.* **51**, 561 (1955).
46. Bain, C. D., Davies, P. B. & Ward, R. N. In-Situ Sum-Frequency Spectroscopy of Sodium Dodecyl Sulfate and Dodecanol Coadsorbed at a Hydrophobic Surface. *Langmuir* **10**, 2060–2063 (1994).
47. Fainerman, V. B. *et al.* Surface tension isotherms, adsorption dynamics and dilational visco-elasticity of sodium dodecyl sulphate solutions. *Colloids and Surfaces A: Physicochemical and Engineering Aspects* **354**, 8–15 (2010).
48. Borukhov, I., Andelman, D. & Orland, H. Steric Effects in Electrolytes: A Modified Poisson-Boltzmann Equation. *Phys. Rev. Lett.* **79**, 435–438 (1997).
49. Kilic, M. S., Bazant, M. Z. & Ajdari, A. Steric effects in the dynamics of electrolytes at large applied voltages. I. Double-layer charging. *Phys. Rev. E* **75**, 021502 (2007).
50. Zdziennicka, A., Szymczyk, K., Krawczyk, J. & Jańczuk, B. Activity and thermodynamic parameters of some surfactants adsorption at the water–air interface. *Fluid Phase Equilibria* **318**, 25–33 (2012).

## Data availability

The authors declare that the data supporting the findings of this study are available in the paper and source data files. Should any raw data files be needed in another format, they are available from the corresponding author on reasonable request. Source data are provided with this paper.

## Acknowledgements

This work was supported by the National Natural Science Foundation of China (Grant Nos. 12125403, 12293053, 12221004, and 12250002), the Shanghai Municipal Science and Technology (Grant Nos. 23dz2260100 and 23JC1400400) and the National Key Research and Development Program of China (Grant Nos. 2021YFA1400503 and 2021YFA1400202). We thank E. Tyrode and Y.-C. Wen for advice on the SDS purification.

## Author contributions

Y.-Y.P. and C.-S.T. devised the project. Y.-Y.P. performed the sample preparation and PS-SFVS experiments. Y.-Y.P., F.G. and C.-S.T. analysed the data. Y.-Y.P. and C.-S.T. drafted the manuscript and all authors contributed to the final version.

## Competing interests

The authors declare no competing interests.

# Extended Data

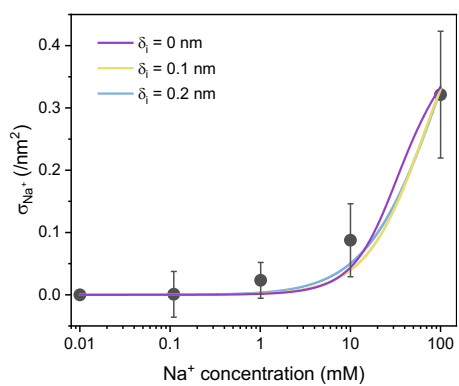

**Extended Data Fig. 1 | The effect of inner Helmholtz layer thickness.** The best-fitted curves of the black scatters with different $\delta_i = 0,\ 0.1,\ 0.2\ nm$ and corresponding $\Delta G_{Na^+} = 6.61,\ 4.96,\ 2.92\ kJ/mol$, respectively. The black scatters are deduced from the SFVS measurements (Fig. 3 in main text).

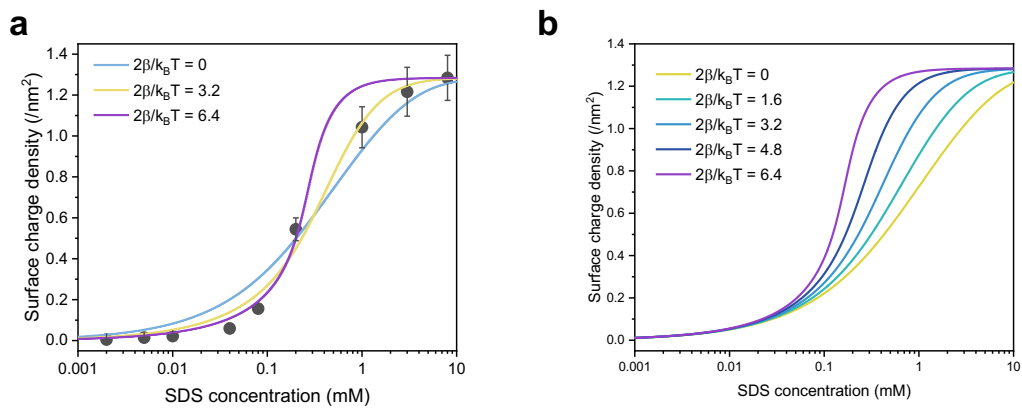

**Extended Data Fig. 2 | The effect of Frumkin interaction parameter. a,** The best-fitted curves of the black scatters with different $2\beta/k_BT = 0, 3.2, 6.4$ and corresponding $\Delta G_{DS^-} = -42.71,\ -38.97,\ -36.60\ kJ/mol$, respectively. The black scatters are deduced from the SFVS measurements (Fig. 2 in main text). **b,** The simulation results of different values of $2\beta/k_BT$ with fixed $\Delta G_{DS^-} = -38.97\ kJ/mol$.

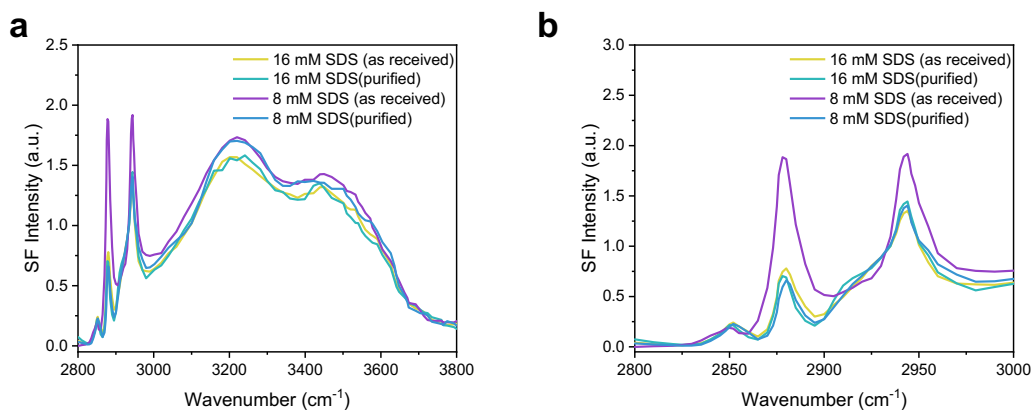

**Extended Data Fig. 3 | Characterization of SDS purification results. a,** SF spectra at air/water interface of 8 mM and 16 mM SDS solution, with and without further purification. **b,** The magnified view of the 2800-3000 cm$^{-1}$ region in **(a)**.

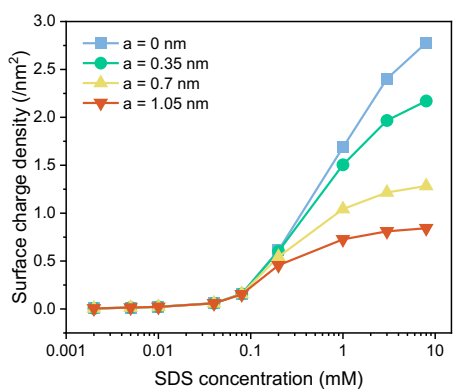

**Extended Data Fig. 4 | Simulations by modified Gouy-Chapman model.** Surface charge densities with different bulk SDS concentrations calculated by different $a$ values using modified Poisson-Boltzmann equation, which degenerates to Poisson-Boltzmann equation when $a = 0\ nm$.

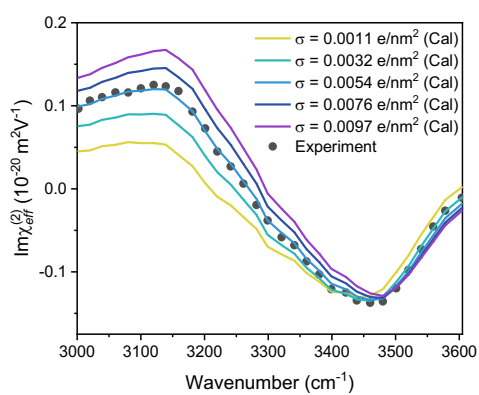

**Extended Data Fig. 5 | Quantitative analysis of SF spectrum.** Experimental spectrum (scatters) and calculated spectra (curves) of 0.002 mM SDS solution for different values of $\sigma$.

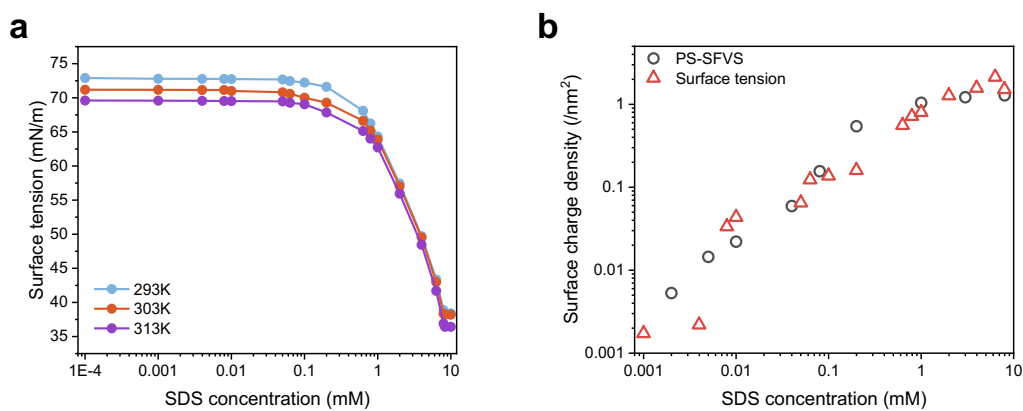

**Extended Data Fig. 6 | Comparison of PS-SFVS and the surface tension measurements. a,** Measured surface tension[50] with various bulk SDS concentrations at 293K, 303K and 313K. **b,** Results for the surface charge density with different bulk SDS concentrations at room temperature, deduced from PS-SFVS and the surface tension measurements. At low SDS concentrations, due to minimal changes in surface tension, surface tension measurements have larger errors.

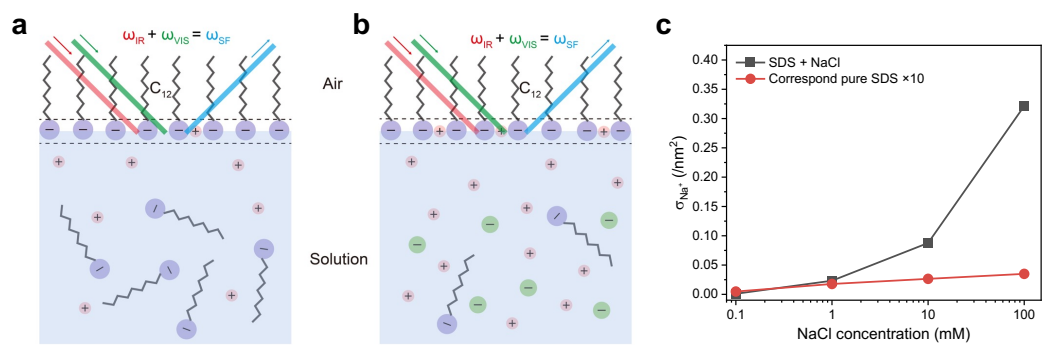

**Extended Data Fig. 7 | Surface Na$^+$ adsorption in pure SDS solution. a,** Illustration of a charged interfacial structure of pure SDS solution probed by PS-SFVS, pink circles represent Na$^+$ and purple circles with tails represent DS$^-$. **b,** Illustration of a charged interfacial structure of SDS/NaCl solution probed by PS-SFVS, green circles represent Cl$^-$. **c,** Surface charge densities of adsorbed Na$^+$ in SDS/NaCl solutions and the corresponding pure SDS solutions. Pure SDS values are scaled ×10 for clarity.